\documentclass[reprint,superscriptaddress,aps,prl]{revtex4-1}
\usepackage{amsmath,epsfig,amssymb,subfigure,bm,dsfont}
\usepackage{lipsum} 
\usepackage{color}
\usepackage{graphicx}
\usepackage{dcolumn}
\usepackage{bm}
\usepackage{bbold}
\usepackage{upgreek}
\usepackage{amsfonts,color}
\usepackage{amsmath}
\usepackage[unicode=true, breaklinks=false, pdfborder={0 0 1}, backref=false, colorlinks=true, linkcolor=blue, urlcolor=blue, citecolor=blue]{hyperref}

\newcounter{lastnote}

\begin{document}
\title{Orbital Pumping in Ferrimagnetic Insulators}

\author{Hanchen Wang}
\email{hanchen.wang@mat.ethz.ch}
\affiliation{%
Laboratory for Magnetism and Interface Physics, Department of Materials, ETH Zurich, Zurich 8093, Switzerland
}%

\author{Min-Gu Kang}
\affiliation{%
Laboratory for Magnetism and Interface Physics, Department of Materials, ETH Zurich, Zurich 8093, Switzerland
}%

\author{Davit Petrosyan}
\affiliation{%
Laboratory for Magnetism and Interface Physics, Department of Materials, ETH Zurich, Zurich 8093, Switzerland
}%

\author{Shilei Ding}
\affiliation{%
Laboratory for Magnetism and Interface Physics, Department of Materials, ETH Zurich, Zurich 8093, Switzerland
}%

\author{Richard Schlitz}
\affiliation{%
Laboratory for Magnetism and Interface Physics, Department of Materials, ETH Zurich, Zurich 8093, Switzerland
}%

\author{Lauren J. Riddiford}
\affiliation{%
Laboratory for Mesoscopic Systems, Department of Materials, ETH Zurich, 8093, Zurich, Switzerland
}%
\affiliation{%
PSI Center for Neutron and Muon Sciences, 5232 Villigen PSI, Switzerland
}%
\author{William Legrand}
\email{william.legrand@neel.cnrs.fr}
\affiliation{%
Laboratory for Magnetism and Interface Physics, Department of Materials, ETH Zurich, Zurich 8093, Switzerland
}%
\author{Pietro Gambardella}
\email{pietro.gambardella@mat.ethz.ch}
\affiliation{%
Laboratory for Magnetism and Interface Physics, Department of Materials, ETH Zurich, Zurich 8093, Switzerland
}%
\date{\today}

\begin{abstract} 
We report the detection of pure orbital currents generated by both coherent and thermal magnons in the magnetic insulator Bi-doped yttrium iron garnet (BiYIG). The pumping of orbital and spin currents is jointly investigated in nano-devices made of naturally oxidized Cu, pure Cu, Pt, and Cr. The absence of charge conduction in BiYIG and the negligible spin-to-charge conversion of oxidized Cu allows us to disambiguate the orbital current contribution. Comparative measurements on YIG and BiYIG show that the origin of the orbital pumping in BiYIG/oxidized Cu is the dynamics of the orbital magnetization in the magnetic insulator. In Cr, the pumping signal is dominated by the negative spin Hall effect rather than the positive orbital Hall effect, indicating that orbital currents represent a minority of the total angular momentum current pumped from the magnetic insulator. Our results also evidence that improving the interfacial transparency significantly enhances pumping efficiencies not only for spin, but also for orbital currents.
\end{abstract}

\maketitle
Harnessing and understanding non-equilibrium angular momenta of electrons in solids has attracted increasing attention over the past two decades~\cite{manchon2019current,chumak2015magnon,demidov2017magnetization,bauer2012spin,maekawa2023spin}. Most studies have focused on the spin degree of freedom because spin current can easily be generated by charge currents, through the spin Hall effect~\cite{sinova2015spin} and the Rashba-Edelstein effect~\cite{bihlmayer2022rashba}. The exploration of spin currents at a fundamental level primarily revolves around the investigation of spin torques~\cite{manchon2019current,ralph2008spin} and its counterpart, spin pumping~\cite{tserkovnyak2005nonlocal,harder2016electrical,yang2018FMR}. These phenomena have not only provided valuable insights for generating pure spin currents~\cite{tombros2007electronic,kajiwara2010transmission} and controlling magnetization~\cite{yang2008giant,miron2011perpendicular,liu2012spin,caretta2018fast,grimaldi2020single} but have also paved the way for practical applications of spintronics~\cite{torrejon2017neuromorphic,krizakova2022spin,sun2022spin,blasing2020magnetic,luo2020current}. 

In contrast, the orbital degree of freedom has largely been overlooked, as it was commonly assumed to be quenched in solids because of strong crystal field potentials and electron delocalization. Recent theoretical~\cite{bernevig2005orbitronics,tanaka2008intrinsic,kontani2009giant,go2018intrinsic,han2022orbital,salemi2022firstprinciples,pezo2024theory} and experimental~\cite{choi2023orbital,lyalin2023magneto,sala2023orbital,ding2022observation} studies have however unveiled that a net flow of orbital momenta can be induced transverse to an electric field out of equilibrium, through momentum-space orbital texture effects. Unlike spin currents, orbital currents cannot directly interact with a magnetization through exchange, so that spin-orbit coupling (SOC) is needed to convert orbital accumulation into spin accumulation and create a torque~\cite{go2020orbital,go2023longrange, gao2018intrinsic,lee2021orbital,ding2020harnessing,lee2021efficient,sala2022giant,hayashi2023observation,nikolaev2024large}. The large orbital Hall conductivity and the tunable SOC, provided by the ferromagnet itself or by a metallic layer nearby the ferromagnet~\cite{ding2020harnessing,lee2021efficient,sala2022giant,ding2024orbital,nikolaev2024large}, lead to strong orbital torques that can be harnessed for electrical manipulation of the magnetization~\cite{zheng2020magnetization,yang2024orbital}. 

These advances in orbitronics have sparked interest into the inverse phenomenon of orbital torque, that is, how magnetization dynamics can generate charge currents via orbital pumping~\cite{go2023orbital,han2023theory}. Recently, the occurrence of orbital pumping has been demonstrated either directly from a ferromagnet~\cite{hayashi2023pump,elhamdi2023observation}, or indirectly by the conversion into orbital current of a pumped spin current~\cite{santos2024exploring,santos2023inverse,santos2024bulk}. These experimental efforts have predominantly focused on metallic magnetic systems, or insulators incorporating additional nonmagnetic metal spacers with strong SOC to facilitate the conversion of spin into orbital pumped currents~\cite{hayashi2023pump,elhamdi2023observation,santos2024exploring,santos2023inverse,santos2024bulk,mendoza2024efficient}. In these systems, however, the presence of large SOC complicates the distinction between spin and orbital transport. Replacing magnetic metals by magnetic insulators as the pumping source modifies the exchange interaction in addition to its conducting character~\cite{go2023orbital}. It also prevents any flow of electrons that would cause out-of-equilibrium orbital or spin diffusion in the ferromagnet itself and could lead to additional spin-orbital inter-conversion through SOC. Therefore, the exploration of orbital currents directly generated from the magnetization dynamics of an insulator, and without a stage of spin-orbital inter-conversion, is expected to provide crucial insight into the fundamental principles of orbital pumping.

In this Letter, we experimentally investigate the orbital and spin pumping efficiencies from ferrimagnetic insulator Bi-doped YIG (BiYIG) into naturally oxidized Cu (denoted by Cu*), pure Cu, Cr and Pt, without inserting converter layers. A significant orbital pumping from BiYIG into Cu* is detected, owing to the interfacial orbital Rashba-Edelstein effect at the Cu/CuO$_x$ interface~\cite{gao2018intrinsic,kim2023oxide,kageyama2019spin,go2021orbital,kim2021nontrivial,kim2023oxide,krishnia2024quantifying,ding2024orbital,ding2024mitigation,santos2024exploring,santos2023inverse,santos2024bulk,mendoza2024efficient}. We further address the origin of this orbital pumping, by replacing BiYIG with YIG. Since in ferrimagnetic insulators there are no conducting electrons to activate the conversion of a spin current into an orbital current, we assign the orbital current generation to the orbital component of magnetization dynamics itself. This demonstrates the occurrence of direct orbital pumping. In the case of pumping into Cr, the signal features a competition between orbital- and spin-charge conversion effects, which we establish to be dominated by spin currents. We further demonstrate that Ar$^+$ etching of BiYIG/Cu interfaces enhances interface transparency not only for spins, but also for orbitals, reinforcing the process of direct orbital pumping.

\begin{figure}
\hspace{-0.5cm}\includegraphics[width=90mm]{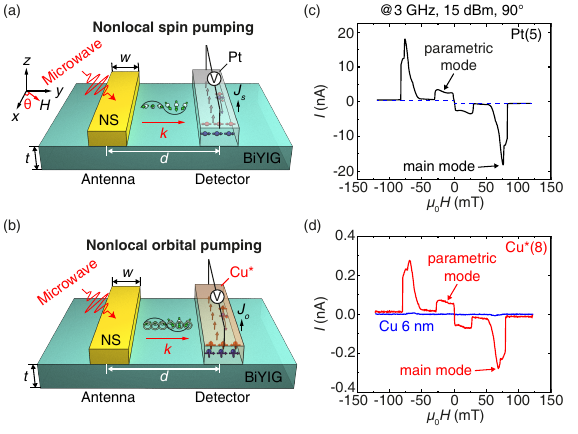}
\caption{Illustrative diagrams of the nonlocal spin (a) and orbital (b) pumping device. Field-dependent converted currents with an excitation frequency of 3.0~GHz, measured on Pt(5) (c), Cu*(8) and pure Cu(6) detectors (d). The field is applied along $y$, and the power is 15~dBm. }
\label{fig1}
\end{figure}

For most of the measurements that follow, a low-damping BiYIG thin film is used, epitaxially grown on a (111)-oriented yttrium scandium gallium garnet (YSGG) substrate by radio-frequency magnetron sputtering at high temperature, with thickness $t=20$~nm. More details are provided in Refs.~\cite{legrand2024lattice,SI}. While the electrical signal generated by orbital pumping can be improved by the deposition of additional converter layers with strong SOC~\cite{santos2024exploring,santos2023inverse,santos2024bulk}, here we intend to avoid converter layers, which requires better experimental sensitivity. We therefore choose to investigate spin and orbital pumping relying on a broad-wavevector nonlocal pumping technique~\cite{wang2024broad}, as presented in Figs.~\ref{fig1}(a) and~\ref{fig1}(b). The large magnon populations generated by the localized microwave field and thermal excitation in nano-devices considerably enhance signals from coherent pumping and thermoelectric effects, respectively, allowing us to detect small conversion effects with a high signal-to-noise ratio~\cite{SI}. 

All the devices with a BiYIG interface presented below are fabricated on a single chip to ensure their similar crystalline quality and identical magnon mode linewidth, for a consistent comparison of pumping efficiencies. Each device features a 420~nm-wide nano-stripe (NS) antenna made of Ti/Au, directly deposited on BiYIG to excite coherent propagating magnons with a broad-wavevector content and incoherent thermal magnons. To detect the spin or orbital currents pumped from magnon excitations, another NS, acting as a detector via spin- or orbital-charge conversion, is deposited by DC sputtering and lift-off, located at $d=2$~$\upmu$m from the NS antenna. This NS detector is made of either Pt, Cr, Cu, or Cu* to investigate spin and orbital pumping in these different metals~\cite{SI}. A 10-nm-thick SiN$_{\rm x}$ capping layer was employed to protect the non-oxidized Cu. The topography of a typical device is presented in~\cite{SI}.

The excitation frequency in the NS antenna is fixed at 3~GHz, with a microwave power $P$ up to 15~dBm, and the external magnetic field is applied along the $y$ direction to achieve a maximal detection efficiency. The propagating magnons excited by the NS antenna pump spin and orbital currents into the metallic NS detector, with a polarization aligned with the static magnetization along $y$. These pumped angular momentum currents are converted into a transverse open-circuit voltage, the magnitude of which depends on the relative spin- and orbital-charge conversion efficiencies in the metallic detector~\cite{mosendz2010quantifying}. As we detail later on, differences in spin and orbital currents pumped in different detector materials may also arise from different transparencies of the insulator/metal interface to transmit angular momentum, known as spin or orbital mixing conductance. 

To illustrate the experimental procedure, we begin by comparing two different detectors: Pt, typical for spin-to-charge conversion experiments, and Cu*, which is sensitive to orbital currents. All measured voltages are normalized by detector resistance to be expressed as converted charge currents. Figures~\ref{fig1}(c) and~\ref{fig1}(d) show converted current against external magnetic field for Pt(5) and Cu*(8) detectors, where the numbers in parentheses indicate the layer thickness in nm. Apart from a significant difference in amplitude, the very similar field dependence of the converted currents in Pt(5) and Cu*(8) confirms their common origin in the different magnon modes excited in BiYIG. By comparing Cu*(8) and Cu(6) detectors (Fig.~\ref{fig1}(d)), the negligible signal from non-oxidized Cu indicates that the Cu/CuO$_x$ interface plays a crucial role in detecting a sizeable converted current in Cu*.

The marked peaks around $\pm$70~mT are attributed to propagating magnons with broad-wavevector distribution. Under the large excitation power employed here, the system enters a non-linear regime~\cite{SI}. This is evident as the peaks deviate from a conventional Lorentzian shape broadened by the NS antenna response, displaying instead a sharp step at higher fields due to the non-linear foldover effect~\cite{elyasi2020resources,ando2012spin,wang2023deeply,suhl1960foldover}. Contributions to the converted current due to resonance-induced heating are estimated to be negligible~\cite{SI,tu2017bolometric}.
Additionally, parametric pumping causes a plateau signal near zero field~\cite{sandweg2011spin,sheng2023nonlocal}. A more extensive discussion of this mechanism is provided in~\cite{SI}. Besides coherently excited magnons, Joule heating in the NS antenna excites thermal magnons with no phase or frequency coherence. These thermal magnons are detected by an incoherent pumping process, known as the spin or orbital Seebeck effect~\cite{uchida2010spin}, induced by the out-of-plane temperature gradient beneath the detectors. 
Because the frequency of thermal magnons extends up to the THz range, the applied magnetic field has a negligible effect on the thermal magnon population. The converted current from incoherent magnons thus appears as a flat baseline reversing for opposite field directions, and a step near zero field hidden by the parametric pumping plateaus, see the dashed blue lines in Fig.~\ref{fig1}(c). The reversal with magnetic orientation is a signature of a spin or orbital Seebeck signal. Overall, our broad-wavevector nonlocal pumping technique facilitates the detection of angular momentum currents, concurrently generated by coherent and thermal magnons.

\begin{figure}
\hspace{-0.5cm}\includegraphics[width=90mm]{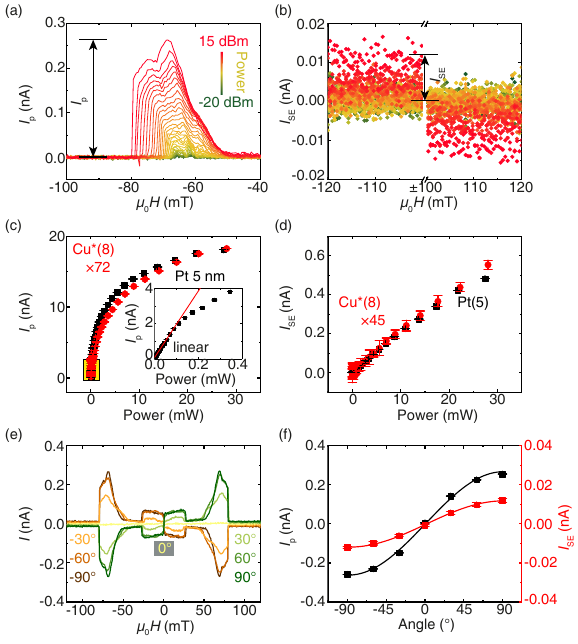}
\caption{Power-dependent converted currents induced by propagating (a) and thermal magnons (b) measured on Cu*(8) device. Amplitudes of coherent (c) and thermal (d) pumping currents as a function of microwave power, comparing Pt(5) and Cu*(8). Inset in (c) highlights the transition from linear to non-linear regime. (e) Field-dependent coherent and thermal converted currents measured in Cu*(8), for different external magnetic field angles relative to the nano-stripe ($P=$~15~dBm). (f) Converted currents from coherent (black) and thermal (red) magnons as a function of field angle, solid lines are sinusoidal fits.}
\label{fig2}
\end{figure}

To fully characterize coherent vs.\ thermal contributions, we further quantify the converted currents in Pt(5) and Cu*(8) detectors under varying excitation powers. Our analysis distinguishes contributions by separating the baseline in the field dependence (thermal, always active) from the peaks (coherent, active only at certain field-frequency conditions). The peaks for Cu*(8) at different $P$ are shown with subtracted baseline in Fig.~\ref{fig2}(a). The corresponding large-field baselines (above $\pm$100~mT to fully exclude contributions from coherent magnons) are shown in Fig.~\ref{fig2}(b). The peak amplitude and averaged baseline for a given field direction define the coherent and incoherent pumping signal $I_{\rm{p}}$ and $I_{\rm{SE}}$, respectively. An identical analysis is performed for Pt(5)~\cite{SI}. The coherent pumping in Cu*(8) and Pt(5) exhibit a consistent nonlinear behavior as a function of $P$, evidenced by scaling $I_{\rm{p}}$ in Cu*(8) by an amplitude difference of a factor about 72 (Fig.~\ref{fig2}(c)). The inset in Fig.~\ref{fig2}(c) focuses on low powers, highlighting a linear regime for $P<$ 0.1~mW. Above this value, a clear deviation from linearity locates the transition to the nonlinear regime (complete characterization in SM~\cite{SI}). By contrast, $I_{\rm{SE}}$ for both Cu* and Pt detectors follow a linear dependence on $P$ up to the largest power of 15~dBm (Fig.~\ref{fig2}(d)). The difference in scaling ratios between Pt and Cu* for $I_{\rm{p}}$ ($\sim72$) and $I_{\rm{SE}}$ ($\sim45$) is expected, because of different thermal conductivities affecting $I_{\rm{SE}}$ in Pt and Cu. The measurement of $I_{\rm{p}}$ and $I_{\rm{SE}}$ for $P=$~15~dBm at different field angles $\theta$, shown in Figs.~\ref{fig2}(e) and~\ref{fig2}(f), exhibit the sinusoidal dependence expected for both coherent and incoherent converted charge currents, validating their interpretation and excluding sizeable spin rectification effects~\cite{SI}. 

For each detector, the dependence on $P$ of the coherent contribution $I_{\rm{p}}$ in the linear regime (of the thermal contribution $I_{\rm SE}$) determines the charge generation efficiency, $\kappa_{\rm{p,SE}}=I_{\rm{p,SE}}/P$, resulting from the sum of spin and orbital processes. In deriving $\kappa_{\rm{p}}$ and $\kappa_{\rm SE}$, the actual power injected in the antenna is carefully evaluated, accounting for the variations in impedance caused by nano-fabrication~\cite{SI}.

\begin{figure}
\hspace{-0.5cm}\includegraphics[width=90mm]{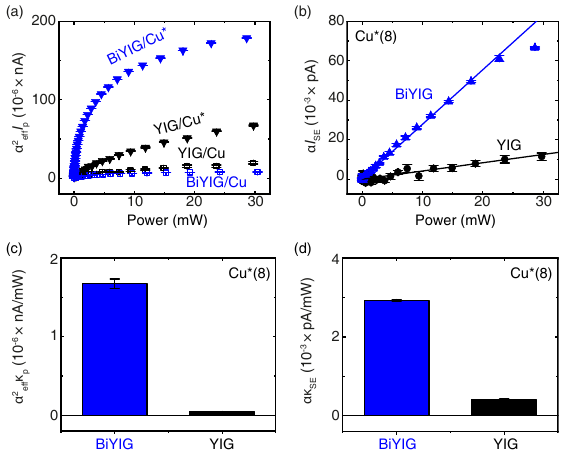}
\caption{Power-dependence of normalized coherent converted currents for BiYIG/Cu*, BiYIG/Cu, YIG/Cu*, and YIG/Cu systems (a) and of thermal converted currents for BiYIG/Cu* and YIG/Cu* (b). Normalized coherent (c) and thermal (d) charge generation efficiencies estimated from the linear regime. The limit power for a linear coherent pumping signal, used for analysis, are found to be 0.07 and 4.00~mW for BiYIG and YIG, respectively. All these devices incorporate Ar$^+$-etched interfaces. }
\label{fig3}
\end{figure}

The origin of the pumping in BiYIG/Cu* has not been investigated before and deserves further attention. The $g$-factor of BiYIG is about 2.03~\cite{SI}, larger than 2 and consistent with the finite orbital magnetization expected of this material~\cite{li2021magneto,rogalev2009element}. Unsubstituted YIG, on the other hand, exhibits $g$-factor about 2.01, consistent with a vanishing orbital magnetization component~\cite{vasili2017direct,cheshire2022absence}. We thus perform comparative measurements with Cu and Cu* detectors on unsubtituted YIG. Due to very small signals, we resort to mild Ar$^{+}$ ion etching of the YIG/Cu and YIG/Cu* interface before deposition to improve their interfacial transparencies~\cite{SI, putter2017impact,yang2019giant}. Accordingly, new BiYIG/Cu and BiYIG/Cu* detectors are fabricated with identically etched garnet/metallic detector interfaces. Since different ferrimagnetic insulators are investigated, their unequal magnetic dissipation influences the magnetization dynamics and needs to be considered in comparing pumping efficiencies in YIG and BiYIG. 
Furthermore, the coherent signal scales proportionally to the square of the precession angle $\phi$, yielding $I_{\rm p} \propto \phi^2 \propto \alpha_{\rm eff}^{-2}$, where $\alpha_{\rm eff}$ represents the effective damping, which accounts for inhomogeneous broadening. In contrast, for the thermal signal generated by high-frequency magnons, the effect of inhomogeneous broadening is negligible, resulting in $I_{\rm SE} \propto \alpha^{-1}$
\cite{SI,rezende2014magnon,chang2017role,iguchi2017measurement,brangham2016thickness,iwasaki2019machine}. Therefore, we multiply
$\kappa_{\rm{p}}$ by $\alpha_{\rm{eff}}^2$ and $\kappa_{\rm SE}$ by $\alpha$ (since in either case, the garnet thickness is smaller than the magnon relaxation length), where $\alpha$ is the Gilbert damping and $\alpha_{\rm{eff}}=\gamma\mu_0 \Delta H/2\omega$ with $\mu_0 \Delta H$ the FMR linewidth~\cite{SI}. As shown in Fig.~\ref{fig3}(a), the sizeable signal detected in BiYIG/Cu* is strongly suppressed in YIG/Cu*, while the signal obtained in BiYIG/Cu and YIG/Cu remains consistently small. Furthermore, a Cu(5)/Pt(5) detector with an identically etched interface exhibits a large signal~\cite{SI}, ensuring that the interfacial transparency in the present YIG or BiYIG/Cu or Cu* devices is substantial. Spin and orbital currents are thus able to escape the garnet/Cu metal interface~\cite{du2014enhancement} and reach the orbital Rashba-Edelstein active Cu/CuO$_x$ interface of Cu*. 

These results lead to two important deductions. Firstly, pumping in YIG generates a spin current with a negligible associated orbital current; secondly, Cu* essentially ensures orbital-charge conversion but no spin-charge conversion. A much larger converted charge current would be present in YIG/Cu* otherwise. The consistent absence of signal in BiYIG/Cu and YIG/Cu rules out the formation of a BiCu layer after etching the interface, which could also generate spin-related signals otherwise. As mentioned in introduction, two possibly competing mechanisms may generate orbital currents in a pumping process: direct generation via orbital magnetization dynamics and conversion of pumped spin currents into orbital currents via SOC of charge carriers. The latter process is not active in insulators due to the absence of conduction electrons. 

The present measurements are therefore consistent with a mechanism in which the charge current in BiYIG/Cu* arises from the direct pumping of orbital currents from BiYIG, followed by orbital-charge conversion within Cu*. In contrast, the spin current generated via spin pumping does not lead to a significant conversion into charge current. The normalized power-dependent signals in Cu*(8) due to thermal magnons in BiYIG and YIG are shown in Fig.~\ref{fig3}(b), and also provide a direct overview of their orbital pumping efficiencies, since Cu*(8) is essentially sensitive to orbital currents. Figures~\ref{fig3}(c) and~\ref{fig3}(d) summarize orbital pumping from coherent magnons (in the linear regime) and thermal magnons, respectively (more details in SM~\cite{SI}). The coherent and thermal signals in each case reflect the same quantities and appear consistent with each other within the accuracy of the normalization method. A more detailed model of coherent and incoherent orbital pumping in materials with different $g$-factors may further improve this consistency.
The vanishing coherent and thermal signals in YIG align with its role as a nearly pure spin source, exhibiting negligible orbital pumping, while Cu* functions effectively as a pure orbital detector. The addition of Bi in YIG is known to enhance its magneto-optical properties~\cite{bi2011onchip,fakhrul2019magnetooptical,li2021magneto}, indicating a sizeable orbital character of the magnetization. The significant coherent and thermal signals observed in BiYIG support the conclusion that orbital pumping in such magnetic insulators arise from their orbital magnetization dynamics ~\cite{rogalev2009element,zhu2023magnetoresistance,zhu2024orbital,emori2024quantifying}, consistently with $g$-factors measured via FMR in YIG and BiYIG~\cite{SI}.

\begin{figure}
\hspace{-0.5cm}\includegraphics[width=90mm]{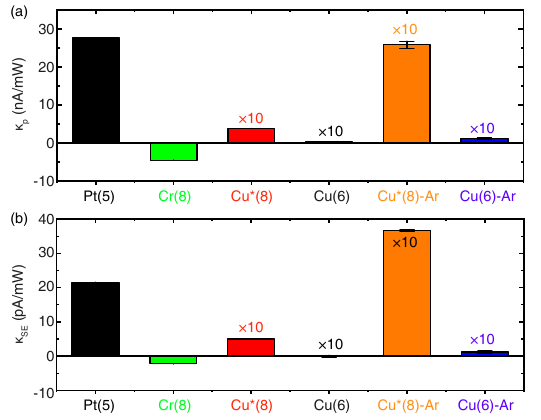}
\caption{Comparison of normalized coherent (a) and thermal (b) charge generation efficiencies in different metallic stacks with four different kinds of BiYIG/metal interfaces: BiYIG/Pt, BiYIG/Cr, BiYIG/Cu, and pre-etched BiYIG/Cu. The efficiencies are both extracted from the linear regimes~\cite{SI}.  }
\label{fig4}
\end{figure}

Next, we introduce the detection of mixed spin and orbital currents in Cr, which has sizeable spin and orbital Hall conductivities but of opposite signs~\cite{salemi2022firstprinciples,du2014systematic, sala2022giant,lee2021efficient,jo2018gigantic}. From the sign of the converted current, we can distinguish whether a spin or orbital process dominates. This distinction is challenging in metallic ferromagnetic systems, where an intrinsic SOC always couples spin and orbital currents. The inverse orbital Hall effect in Cr is expected to convert the orbital pumping current into a positive signal like in Pt. The signal in Fig.~\ref{fig4}(b), negative instead, reveals that spin pumping dominates over orbital pumping in BiYIG, consistent with the negative spin Hall conductivity in Cr.

As appeared above, the pumping signals can be experimentally improved using Ar$^+$ etching before metal deposition. Compared to the unprocessed BiYIG/Cu interface, both coherent and thermal pumping efficiencies are enhanced, roughly by one order of magnitude, for either Cu*(8) (Fig.~\ref{fig4}), with an orbital-charge conversion, or Cu(5)/Pt(5) \cite{SI}, with both spin-charge and orbital-charge conversion. The successful improvement of pumping signals for Cu*(8) shows that, akin to surface treatments that reinforce interfacial spin transparency~\cite{putter2017impact,jungfleisch2013improvement}, Ar$^+$ etching is able to improve the interfacial orbital transparency~\cite{putter2017impact,hayashi2023pump,lyalin2024interface}.

In summary, we experimentally observe sizeable orbital currents pumped from coherent and thermal magnons in BiYIG, detected in naturally oxidized Cu via the inverse orbital Rashba-Edelstein effect. Comparative measurements with YIG indicate that this orbital pumping is related to the dynamical orbital magnetization component of BiYIG. Using Cr as a detector sensitive to both spin and orbital currents with opposite conversion factors, the observed sign of the charge current in BiYIG/Cr shows that spin currents dominate the pumping process. This observation aligns with the predominantly spin-like character of the magnetization dynamics. Additionally, we show that the interfacial orbital transparency can be enhanced through Ar$^{+}$ etching of the BiYIG/Cu interfaces, significantly reinforcing the amplitude of the orbital pumping. These results provide a direct and consistent validation of the theoretical concepts of orbital pumping, and open a promising avenue for investigating pure orbital currents, enabling their clear differentiation from spin currents in non-equilibrium angular momentum transport. Additionally, exploring these mechanisms in antiferromagnets~\cite{han2023coherent}, which host multiple magnon modes and may exhibit orbital magnetization, represents an exciting frontier for novel spin-orbitronic devices.

\begin{acknowledgments}

We thank Gerrit E. W. Bauer for helpful discussions about the normalization of the Seebeck effect. This research was supported by the Swiss National Science Foundation (Grant No. 200020-200465). H.W. acknowledges the support of the China Scholarship Council (CSC, Grant No. 202206020091). M.-G.K. acknowledges support from the Basic Science Research Program through the National Research Foundation of Korea (Grant No. 2022R1A6A3A03053958). L.J.R. acknowledges support from the ETH Zurich Postdoctoral Fellowship Program (22-2 FEL-006). W.L. acknowledges support from the ETH Zurich Postdoctoral Fellowship Program (21-1 FEL-48).

\textit{Note added.}$-$During the revision of this manuscript, we learned of recent experimental evidence showing that orbital currents can be generated from orbital dynamics in the antiferromagnetic insulator $\alpha$-Fe$_2$O$_3$ using terahertz (THz) emission spectroscopy~\cite{huang2024orbital}.

\textit{Data availability}$-$The supporting data for this article are openly available from the ETH research Collection~\cite{ETH}.

\end{acknowledgments}
\bibliographystyle{unsrt}

\end{document}